# Towards Efficient OpenMP Strategies for Non-Uniform Architectures

Oussama Tahan
Department of Computer and Communication Engineering
Lebanese International University (LIU), Beirut, Lebanon
oussama.tahan@liu.edu.lb

*Abstract*—**Parallel processing is considered as todays and future trend for improving performance of computers. Computing devices ranging from small embedded systems to big clusters of computers rely on parallelizing applications to reduce execution time. Many of current computing systems rely on Non-Uniform Memory Access (NUMA) based processors architectures. In these architectures, analyzing and considering the non-uniformity is of high importance for improving scalability of systems. In this paper, we analyze and develop a NUMA based approach for the OpenMP parallel programming model. Our technique applies a smart threads allocation method and an advanced tasks scheduling strategy for reducing remote memory accesses and consequently their extra time consumption. We implemented our approach within the NANOS runtime system. A set of tests was conducted using the BOTS benchmarks and results showed the capacity of our technique in improving the performance of OpenMP applications especially those dealing with a large amount of data.**

## I. Introduction

Since decades, the underlying architectures of computers have been constantly evolving and the new technology advances are still providing microprocessors with better capacities. The increasing complexity of applications and the higher number of functionalities microprocessors need to ensure have led computer designers to improve drastically these architectures in order to obtain better performance.

In the last few years, processors manufacturers abandoned efforts aiming to increase clock speed of modern processors and decided to study different approaches. These methods are mainly based on exploiting parallelism in MIMD (Multiple Instruction Multiple Data) architectures like multi-core and many-core based ones.

This strategy aims to achieve better computing performance while reducing or keeping the same level of power consumption.

Today, parallel based computing is the dominating used approach for obtaining high performance in current systems. With the important trend shift of processor development from uni-core to multi and many-core, designers are encouraged to parallelize applications. This has urged the need for parallel programming tools and efficient programming models in order to deal with growing applications complexity. Proper runtime systems are also needed to execute parallel applications on existing cores. OpenMP or Open Multi-Processing is a well-established programming model used to efficiently code parallel applications on multi-core architectures. It is a popular standard commonly employed today in several domains due to its portability, flexibility and ease-of-use [1, 2, 3].

One of the challenges facing OpenMP and many other parallel models is their capability of handling parallel computer architectures with distributed memory. These architectures which are also named NUMA (Non-Uniform Memory Access), form a challenge for the programming models and runtime systems that need to effectively distribute execution load on available resources.

The need for an effective load distribution becomes of great importance for applications and benchmarks dealing with high number of tasks and data. Executing these applications on NUMA systems without considering hardware architecture characteristics, has major impact on performance and produces significant slowdowns.

In this paper, we present our work consisting of developing and implementing a new extension to an open source and flexible OpenMP runtime system, so that OpenMP developers can execute efficiently parallel applications on NUMA systems. Our system is based on a smart thread to core allocation and it introduces new task based schedulers to minimize accesses of threads to distant memory locations.

To present our contribution, this paper is structured as follows: In the following section we present in some details the basic characteristics of NUMA architectures and we illustrate the reason behind our study. In Section III, we show some work related to ours while in Section IV, we introduce our own threads management and allocation system. In Section V, we show the results of our approach and we demonstrate the obtained improvements. In Section VI, we present the task schedulers we developed for taking into consideration NUMA based systems and we illustrate the additional obtained improvements. Finally, in Section VII we conclude and present our future work.

## II. NUMA ARCHITECTURES

These systems are a combination of both UMA (Uniform Memory Access) [4] and Distributed Memory architectures [5]. In NUMA based systems, memory is physically distributed on processors and hence, each processor is directly connected to a part of the memory to form a node. However, unlike distributed architectures and similar to UMA, processors share a global address space. Hence, reading or writing to a variable on a distant node does not have to go through explicit data fetch or write requests to a processor located on that node.

These architectures are called NUMA or Non Uniform Memory Access because access time to a variable in memory is not uniform. It may vary depending on location of the processor and the variable in memory; accessing data in local memory is faster than accessing distant one. In addition, time spent to access remote memory locations is not always identical. It depends on distance separating the core and the NUMA memory node to which it is trying to access. In most NUMA architectures, several distances may exist and hence, a system may have several NUMA factors. NUMA factor is the ratio between latency for accessing data from local memory and for accessing distant memory location. In this domain, the most popular metric to determine distance between two NUMA nodes is hop. From now on we will refer to distances between nodes as number of hops. A simple structure of a NUMA architecture is presented in Figure 1.

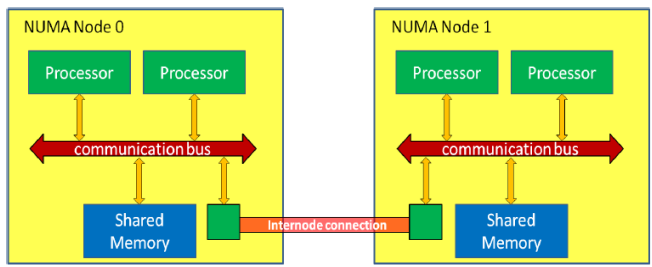

Figure 1 Non Uniform Memory Access (NUMA) computer architecture

NUMA systems were conceived to overcome issues of both UMA and distributed memory architectures by decreasing the rate and level of bus contention when concurrent memory accesses are requested and by reducing burden for programmers when writing parallel applications. However, applications running on these systems must be carefully parallelized to reduce overheads of high memory access latencies.

Many NUMA based systems are employed today in embedded systems, servers and computing farms domains. We can mention some examples of NUMA systems like AMD Opteron based platforms [6], computer architectures based on Intel Xeon and Intel Nehalem processors family [7], AMD Magny-Cours based architectures [8], IBM Power6 [9], STI Cell [10], SuperH [11], Raw Microprocessor based system [12], TilePro64 [13] and many other architectures.

Lately, this particular type of parallel multi-core architectures initiated new interesting area of research. This is to analyze different approaches, for efficiently exploiting NUMA systems and for providing applications with a way to obtain better performance and faster execution. In parallel applications, work balancing and communication latencies affect dramatically execution time of programs. Therefore, sharing work between cores and properly handling data located in memory, are both essential to efficiently execute parallel programs on NUMA platforms.

Discharging programmers from this difficult and time consuming task is highly desirable. A parallel application adapted to a specific NUMA multi-core system may offer good performance. However, the same application may not necessarily present good execution time on other systems since they do not have the same NUMA architecture. For this reason, implementing NUMA aware programming models, compilers and runtime systems is essential to reduce programmers efforts in coding NUMA dedicated and portable parallel applications for these particular platforms.

## III. RELATED WORK

In this research area, very few techniques were studied to implement NUMA aware OpenMP environment. Some of them are based on thread-centric OpenMP while others are based on task-centric model with task scheduling strategies.

### A. Existing NUMA based Thread-Centric OpenMP Runtime

ForestGOMP is an OpenMP runtime system that captures the structure of parallelism in an OpenMP application and gathers a team of threads issued from the same parallel region into one same group called bubble [14]. A hardware architecture detection called hwloc was implemented to detect different characteristics and features of underlying hardware. They implemented in this runtime two different thread based schedulers. One focuses on scheduling and migrating threads that constitute one bubble in a compact way in order to benefit from caches shared between cores. Another scheduler is based on a joint allocation of threads and data to take into consideration memory affinities defined at user level through specific interfaces. This runtime system schedules threads of the same nested parallel region on the same chip through work-stealing based scheduler. When inter chip work stealing takes place, stolen threads are those with minimum coupled memory; data also migrates with stolen thread. This runtime system showed good improvement in performance for different applications. However, their approach ignores new task-centric OpenMP and focuses on old thread-centric programming model. In this latter, scheduling strategy is completely related to threads and applications use a great number of nested parallel regions. Those regions turned out to be very costly for OpenMP applications. In addition, migrating a thread and memory to which it is coupled is a time costly operation causing loss in performance when large memory is used.

*B. Existing NUMA based OpenMP Task Schedulers*

Other OpenMP implementations were also studied recently in order to take into consideration increasing complexity of multi-core systems and different NUMA architectures. These implementations were intended to be task-centric OpenMP friendly in order to be efficient for future's task based OpenMP applications. In these runtime implementations different task level scheduling strategies were investigated to dynamically schedule tasks in a NUMA profitable manner. In [15], an OpenMP tasks work-stealing strategy called LOCAWR has been implemented within the Nanos runtime system. LOCAWR relies on extensions added to the task creation construct of OpenMP. Using these extensions, the programmer is able to specify start address, scope and size of shared data elements. At task creation point, the parent task analyzes this information to detect to which worker thread this new task should be attached to preserve locality. When this task is submitted, it will be added to the task queue of worker thread for which it has affinity. When a thread is idle, it attempts to pick a task from its own task pool. If empty, it tries to steal a task from one hop distant worker thread. They evaluated their approach on few benchmarks running on tile-based multi-core architecture. They obtained low performance improvement for these benchmarks when using this scheduler while both Nanos cilk-based and dynamic breadth-first schedulers gave better execution time. LOCAWR did not provide sufficient enhancement because it suffers from load imbalance in some benchmarks.

In another approach, Olivier et al. [16] introduced a hierarchical task scheduling strategy called MTS (Multi-Threaded Shepherds). They implemented an OpenMP task scheduler as an extension to existing scheduling strategies of Qthreads library [17] and they compiled OpenMP applications using ROSE compiler [18]. In their hierarchical approach, they create a shepherd per chip, thus one shepherd for all cores of the same chip, considering that these cores share a common L3 cache memory. Cores located on the same chip share a common Last In First Out task queue. Worker threads follow a depth-first scheduling strategy in order to preserve locality. When a queue is empty, one worker tries to steal tasks from a randomly chosen distant queue. Their approach showed good performance on 4-socket Intel Nehalem architecture with 32 cores in total, on 2-socket / 4-chip AMD Magny Cours architecture and on SGI Altix system. However, improvements in execution time were mainly for non-data intensive benchmarks. Other benchmarks did not show satisfying performance enhancement especially on SGI Altix system where NUMA nodes are more than one hop distant away.

## IV. INTRODUCING NUMA AWARE THREADS ALLOCATION TO NANOS

Previous approaches lack an important aspect influencing performance of applications running on NUMA systems. This missing feature relies on taking into consideration underlying hardware architecture to apply good threads to cores allocation and binding.

In our work, we look into NUMA system architecture on which we are running the parallel application. More precisely, we consider positions of cores, nodes and their distances to correctly bind created threads to available cores.

This approach allows us to enhance thread to core allocation in a way that number and distance of remote memory accesses may be reduced. For this purpose, we added to Nanos a first extension aiming to explore the system's architecture and detect number of existing NUMA nodes, distances between them and number of cores per node. This hardware exploration technique is inspired from hwloc library. To obtain this knowledge about underlying hardware, we used APIs of libNUMA [19] and CPU Affinity [20] Linux libraries that can be used within C code by including respectively both **#include <numa.h>** and **#include <sched.h>**. libNUMA offers a programming interface to non-uniform memory access architectures allowing programmers to obtain different propoerties of the NUMA hardware. In addition it allows them to allocate memory on specific nodes. As for CPU Affinity library, it helps programmers to detect the number of existing cores in a system and it allows them to define threads to cores affinities, thus forcing a thread to run on a specific core.

Gathered information allows us to attribute priorities to different cores of the architecture. We define these priorities at the start of a program during runtime initialization phase. This priority can later be changed if any modification of the hardware has taken place.

In our work, we consider the system and all cores are idle and there are no already running workloads. In a first priority attribution level, we assign high priority to cores of the socket/chip having the largest number of cores attached to the same NUMA memory node. This is an approach compliant with NUMA systems, especially with future heterogeneous architectures where number of cores and type per each processor or node may vary. This priority value drops off with decreasing number of cores per NUMA node. If all nodes have equal number of cores, our technique attributes the same priority for all cores of the system. When binding threads on cores directly connected to same memory and in some cases cache memory, we allow tasks ran by these threads to benefit from shared hot caches and memory locality. Hence, they may have less or no distant memory accesses.

---

Let $\alpha_i$ be the weight of each possible hop distance in the system, with $\alpha_i > \alpha_{i+1}$,
where $i \in H$ and H=[min-numa-distance, ..., max-numa-distance].
If i = max-numa-distance then $\alpha_{i+1} = 0$.
Let $N_i$ be the number of cores at i hops from target core.
Value $V_1$ is computed as follows: $V_1 = \sum_i \alpha_i * N_i$

Figure 2 Priority calculation

On another hand, an additional value may be combined to the first computed core priority. This value may be obtained by analyzing the number of cores located at one and more hops from target core. This additional priority may be acquired by assigning for each hop distance a weight. It is a coefficient number decreasing with growing number of hops. Therefore, the value we add to priority increases with the growing number of close cores and may be computed as presented in Figure 2**Error! Reference source not found.**.

The new value $V_1$ is added to the already found priority giving a new priority level for the targeted core.

In a second step, we find for each core another value that may be added to its already computed priority. This new value is associated with core priorities computed earlier, thus close cores with higher priorities offer higher value. This second step is mainly useful for NUMA architectures where several hop distances may exist [22], in heterogeneous multi-core processors (heterogeneous by design or due to core defects) and in case some cores have already been allocated for other work. The new value may be computed like illustrated in Figure 3.

---

Let $\alpha_i$ be the weight of each possible hop distance in the system, with $\alpha_i > \alpha_{i+1}$
where $i \in H$ and H=[min-numa-distance, . . . , max-numa-distance].
If i = max-numa-distance then $\alpha_{i+1} = 0$.
Let $N_i$ be the number of cores at i hops from target core.
Let $P_{ij}$ be the priority of core with index j where $j \in \{0, \ldots, N_{i-1}\}$ and located at i hops from target core.
Value $V_2$ is computed as follows: $V_2 = \sum_i \sum_j \alpha_i * P_{ij}$

---

Figure 3 New added priority value

The final core priority is the sum of old priority and newly computed value. The algorithm we have implemented is shown in Figure 4 . When priorities of cores are found, the Master thread will set its affinity and binds immediately to the core that obtained highest priority. Whether located on same or different nodes, if several cores have the same priority, Master thread will randomly pick and bind to one of them. When new worker threads are created by Master thread, they will be placed as close as possible to its assigned core. When two or more cores are at equal distance from Master thread's core, new worker thread will be assigned to the one with highest priority. If two or more cores have the same priority, one of them will be selected randomly. Cores are picked and bounded to threads in a way that chosen cores are as close as possible. Hence, we eliminate unnecessary large inter-threads distances.

Detecting hardware architecture and setting priorities at runtime start-up allow us to immediately schedule and bind Master thread to the best core and node. Hence, initialization phase of runtime and allocation of data environments take place on the node closest to all cores. In addition to hardware architecture detection and priority allocation, we added proper modifications to Nanos, allowing master and worker threads runtime related data to be allocated on nodes directly connected to cores on which these threads will be running.

## V. PERFORMANCE EVALUATION OF THREADS ALLOCATION MODEL

To test our approach, we used benchmarks provided within Barcelona OpenMP Tasks Suite version 1.1.2 [21]. We ran the eleven benchmarks on the SunFire X4600 system [22] and we applied six tests for each one of them. For each test, we used one of the task scheduling strategies (Breadth-First, Cilk-based or Work-First) provided within Nanos. In the first set of three tests, we used the original runtime implementation to observe effects of these scheduling policies on each benchmark. These tests are presented within figures of the following subsection as follows:

- **bf-Scheduler** is the basic implementation of Nanos using Breadth-First scheduler;
- **Cilkbased-Scheduler** is the basic implementation of Nanos using Cilk-based scheduler;
- **wf-Scheduler** is the basic implementation of Nanos using Work-First scheduler.

For the three remaining tests, we used the same schedulers but this time, we combined them to the thread priority allocation procedure and NUMA enhancements described in Section IV. These tests are presented as follows:

- **bf-Scheduler-NUMA** is the implementation of Nanos using our NUMA aware modifications and using Breadth-First scheduler;
- **Cilkbased-Scheduler-NUMA** is the implementation of Nanos using our NUMA aware modifications and using Cilk-based scheduler;
- **wf-Scheduler-NUMA** is the implementation of Nanos using our NUMA aware modifications and using Work-First scheduler.

---

```
                    Threads Priority Allocation
 1:  set_priorities (void)
 2:  {
 3:    explore_hw_architecture ();  //explores hardware architecture
 4:    weight := assign_weights (); // weight[i] > weight[i+1]
 5:
 6:    for i := 0 to max_core_id
 7:      my_priority := 0;
 8:      for i := 0 to max_numa_distance do
 9:        my_priority := my_priority + number_of_cores_at_hops [i] * weight[i];
10:      end for
11:    end for
12:
13:  // now we have computed first priority level, we will compute second priority level
14:
15:    for i := 0 to max_core_id do
16:      //for each existing core do
17:      temp_priority := 0;
18:
19:        for k := 0 to max_numa_distance do
20:        //find list of cores located at k hops from corei
21:        list_cores := find_cores_on_hops [k];
22:
23:          for j := 1 to size_of(list_cores) do
24:          //take into consideration previously found priorities
25:          temp_priority += get_old_priority (list_cores[j]) * weight[k];
26:          end for
27:
28:        end for
29:
30:      my_priority += temp_priority;
31:    end for
32:  } //end set_priorities
```

Figure 4 Threads priority allocation algorithm

For each benchmark, we tested Medium or Large inputs sizes and for each test we took the best result obtained out of fifty runs. We should note that for some benchmarks no large inputs size are provided within the suite and therefore we limited the tests to Medium inputs. Following, we show the results obtained for some of the benchmarks.

## A. Benchmarks Results

In Figure 5, we show results obtained for Floorplan benchmark and its different speedups on NUMA architecture. For two and four active cores, all implementations showed almost similar and near linear speedups with minor progress for tests based on Cilk and work-first schedulers. For six or more threads, tests relying on the two work-stealing scheduling policies showed better performance than breadth-first based tests. We obtained maximum speedup using the Cilk-based scheduler with NUMA aware implementation on sixteen cores. Adding NUMA aware extensions to Nanos brought respectively almost 3.18% and 3.14% improvement in performance over tests based on Cilk and work-first schedulers without our added changes.

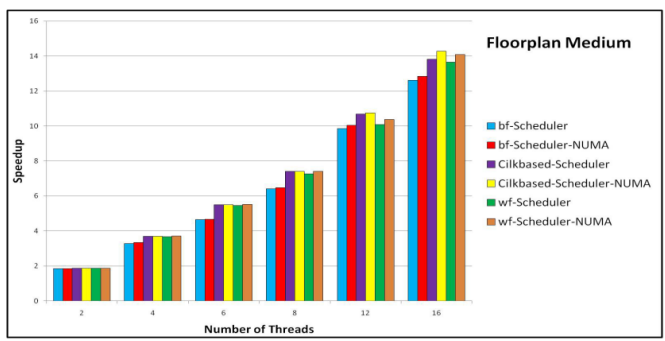

Figure 5 Floorplan benchmark results

SparseLU omp-for version showed results close to those obtained with the single version. Breadth-first scheduler shows the worst performance for more than four cores while work-stealing based tests scale finely with increasing number of threads. We obtained maximum speedup when running the benchmark on sixteen cores with a combination of Cilk-based scheduler and our extensions. Work-first scheduler also leads to good scalability and obtains 13.97x speedup for sixteen cores (see Figure 6). Moreover, combining it with NUMA aware runtime, work-first scheduler yields to 5.24% faster execution time while with the Cilk-based scheduling policy it yields 7.01% improvement.

Running FFT benchmark allows us to clearly identify the weakness of breadth-first scheduler for applications dealing with large amounts of data and tasks. In addition, it shows the importance of our implemented runtime modifications to obtain better performance. FFT is a benchmark that generates around 10M tasks for medium inputs set and 19M tasks for large inputs set. In addition, when using medium inputs, the benchmark's memory utilization is around 6 GBytes while it grows up to 13 GBytes of memory for the large inputs set. When using breadth-first scheduling policy, a thread may pick any task from shared queue of tasks and hence, allows good load balancing. However, a new task selected from shared task pool may probably execute on data not been recently used by the same thread. Hence, data locality is not preserved and local caches are not efficiently exploited.

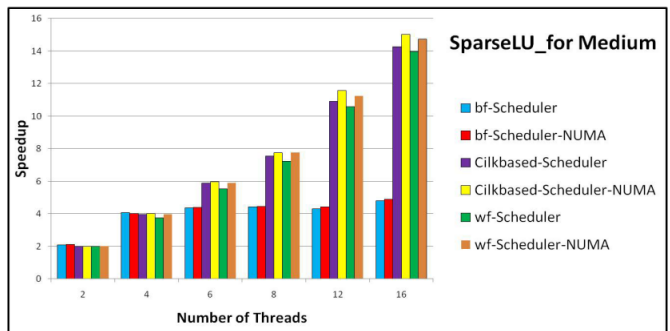

Figure 6 SparseLU_for benchmark speedup

Cilk-based and work-first schedulers solve this issue and preserve data locality by applying depth-first scheduling mechanism. When parent task suspends itself and spawns a new one, it will share data with its child task. Since a copy of this shared data may still be hot in the core's two level caches, number of cache misses may be significantly reduced.

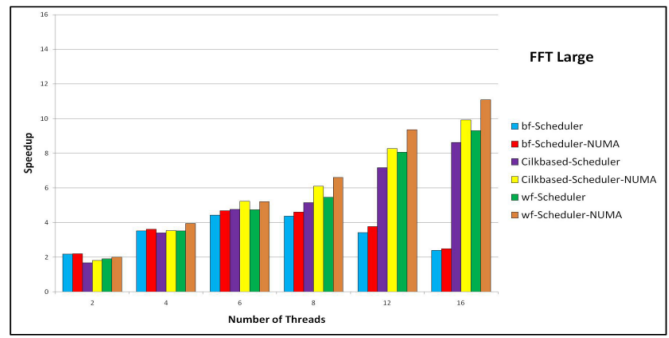

Figure 7 FFT benchmark results

In Figure 7 we notice that breadth-first scheduler increases speedup of the benchmark to a maximum value of 4.43x when using up to six cores. Nevertheless, when running on more than six cores, we detect significant loss in performance with growing number of threads and speedup decreases to reach 2.39x for sixteen cores. Our added NUMA extensions helped improving performance of this scheduler but high queue contentions, important cache misses and high latencies due to distant memory accesses prevent the benchmark from obtaining better speedup.

Work-stealing based implementations, all showed poorer performance than breadth-first when running on two cores. However, they got almost similar execution times when four cores are used except for work-first scheduler combined with our Nanos extensions. This latter runs 8% faster than the combination with the breadth-first scheduler.

When using six or more cores, both Cilk-based and work-first scheduling based implementations obtained good speedups reaching respectively 8.61x and 9.3x for sixteen active cores. Combining these schedulers with our runtime modifications reduced noticeably execution time of FFT where Cilk-based scheduler brought 9.92x speedup to the benchmark. In addition, work-first offered 10.55% faster execution time than Cilk to reach 11.09x speedup.

Like FFT, Strassen requires high memory utilization and consumes around 7 GBytes of memory. Therefore, exploiting data locality is important to benefit from hot caches, reduce distant memory accesses and hence, obtain better improvements in performance.
We notice in Figure 8 that all schedulers showed good performance for different number of active cores. For this benchmark, work-first outperforms other two schedulers for all number of threads and obtains a maximum speedup of about 9.15x when running on sixteen cores. When NUMA adjustments are applied to Nanos, we can clearly see its efficiency in enhancing performance of the application for all used task schedulers. The best two speedups were obtained when sixteen threads were used, for both Cilk-based and work-first scheduling policies combined with our modifications. Both respectively obtain 8.13x and 10.27x speedups.

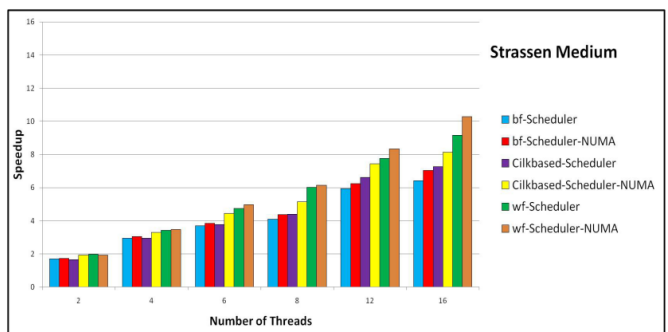

Figure 8 Strassen benchmark speedup

Sort is another benchmark requiring high memory utilization reaching 8.5 GBytes with large input set, with a high number of generated tasks. Figure 9 shows the speedup of Sort benchmark using the different implementations. When two cores are active, work-first obtains the best speedup of 1.86x while Cilk-based scheduler runs 5% slower.

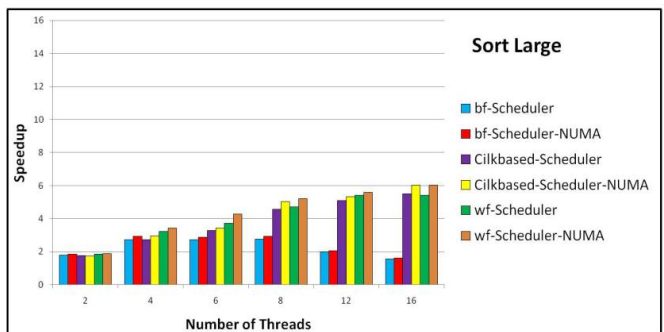

Figure 9 Sort benchmark speedup

For this benchmark, breadth-first shows the worst performance with increasing number of cores because data locality is not preserved and high task queue contentions emerge when using this scheduler. As for remaining implementations based on work-stealing scheduling policies, they both show good improvement in performance. On sixteen active cores, both Cilk-based and work-first schedulers obtain respectively 5.49x and 5.41x speedups. When our modifications are added to Nanos, we obtain respectively 9.17% and 10.06% faster execution time.

For NQueens benchmark, almost all schedulers and implementations brought good speedups for different used threads count. For two and four active cores, all implementations showed similar execution time with linear and super-linear speedups. However, we can see through Figure 10 that when four or more cores are running, NQueens gets its best performance with breadth-first scheduler. This is due to scheduler's effectiveness in balancing the benchmark's workload. NUMA extensions applied to Nanos improved performances of all schedulers. When using breadth-first scheduler alone, the benchmark showed linear and super-linear speedups for almost all number of threads and it obtained on sixteen cores, a maximum speedup of 15.93x over its serial execution time. When adding our NUMA extensions, the benchmark runs 1.35% faster than the basic implementation for sixteen threads.

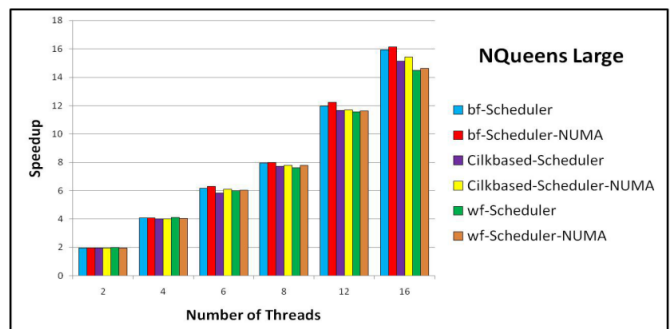

Figure 10 NQueens benchmark speedup

B. *Benchmarks Results Analysis*

After analyzing results obtained in this section, we may clearly notice effectiveness of applying NUMA aware modifications to the runtime system in order to reduce execution time. We noticed an improvement in performance for almost all studied benchmarks. However, this progress in execution time and effects of our approach on speedup differ from a benchmark to another.
We may see through results that our model influences data intensive benchmarks dealing with large amount of data and tasks like FFT, Strassen and Sort. This high efficiency is due to our technique assorted with first-touch NUMA policy.
In modern operating systems (Microsoft Windows, Linux, etc.), a first-touch policy for memory allocation is used when running on NUMA architectures. This policy consists in suspending allocation of physical memory by the operating system to the instant where memory is first accessed through read or write. The operating system will try to allocate pages from local physical node of the CPU having a first read or write memory access to this allocated memory. When not enough free memory is available on that node, allocation falls back to closest nodes [23, 24].

With NUMA aware runtime, master thread runs on the node with highest priority. Hence, it allocates data on the best main and nearby nodes of the architecture (those located as close as possible to all existing cores and to each other). Using basic Nanos implementation, these data are allocated by the operating system on the first node of the architecture. This causes loss in performance for applications running on NUMA based architectures where several core to memory hop distances exist and where memory allocation on the first node is not ideal to obtain good speedups. Sunfire X4600 is an example of these computer architectures and we showed how our method may help improving performance of data and task intensive applications.

## VI. IMPLEMENTING NUMA AWARE TASK SCHEDULERS

Task scheduling is an essential feature in runtime systems. OpenMP task-centric parallel programming model relies on different task scheduling strategies provided within Nanos. This is in order to efficiently schedule generated tasks on underlying parallel processing units, thus obtaining lower execution time.

Today, scheduling strategies provided for OpenMP are not ideal for NUMA based systems. Therefore, applying basic scheduling strategies described earlier (breadth-first, Cilk based and work-first) may not be optimal for applications running on NUMA; especially those applications dealing with high number of tasks and data.

For this purpose, in addition to threads-to-cores allocation technique we presented earlier, we discuss in the following, two NUMA aware task scheduling strategies we implemented in Nanos then we will study their effects on few benchmarks.

### A. DFWSPT Scheduler

In order to take into consideration architectures with varying inter-core distances and memory access latencies, we developed DFWSPT; a depth first scheduler with NUMA aware work stealing. DFWSPT or Depth-First Work-Stealing Priority Threads scheduler is based on a depth-first strategy where each newly spawned task is executed immediately, while its parent is queued in the task pool of the thread on which is has been running. Task queuing takes place in front of the task pool.

At startup, a specific list of threads will be affected for each thread of the team. This list, we call it priority list, contains identification numbers of other threads of the team. These threads are ranked in a priority based manner. Threads assigned to close cores have higher priority than those running on far cores. For example, let us consider that thread #0 is running on core #0, thread #1 on core #1 and thread #2 on core #2 with core #1 located at one hop from core #0 while core #2 is at two hops from core #0.

According to thread #0, thread #1 will have higher priority than thread #2 and hence, it will be placed in the priority list before this latter. If several cores turned out to be at equal distance from target core, threads are placed according to their identification number. Threads with smaller id are placed first. When a thread is idle, it will first look in its own local queue for a task to execute. If there are no tasks to execute, it will start looking for tasks in other queues. However, to do it in a NUMA aware manner, the thread will look into task pools of threads in a priority based order. It will sweep the priority list and checks the pool of tasks of each encountered thread until the end of priority threads array or until a task is picked for execution. Tasks are stolen from the back of the queue of tasks. A general explanation of DFWSPT is shown in Figure 12.

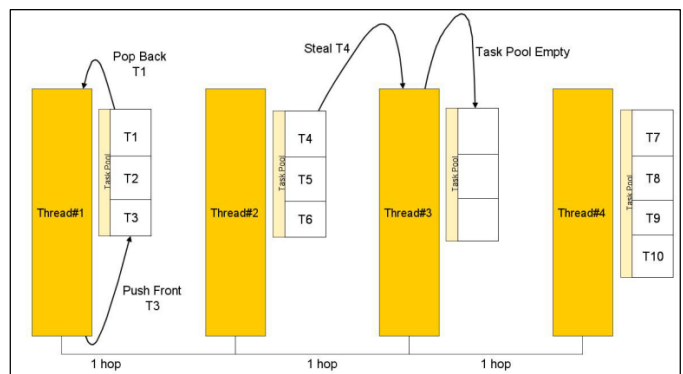
Figure 11 DFWSPT tasks scheduler

### B. DFWSRPT Scheduler

In a similar approach, we implemented this new scheduler within Nanos. Like previously introduced scheduler, DFWSRPT or Depth-First Work-Stealing Random Priority Threads scheduler requires newly created task to execute immediately, while its parent task is queued in the task pool (depth-first mechanism). In addition, a work-stealing mechanism is employed for workload balancing. Like DFWSPT, utilized work-stealing mechanism is NUMA aware and based on priority threads list. However, a victim thread is picked randomly to check its task pool for tasks to execute.

This means that when a thread is idle, it will attempt to steal a task from its own task pool first. If empty, it will try to steal a task from another thread located as close as possible to it. When several threads are at equal distance from the idle thread, whether located on the same NUMA node or not, it will randomly choose its victim thread. Hence, it does not choose its victim thread according to its smaller identification number (see Figure 12). Randomizing thread's selection mechanism may allow applications to avoid contentions that happen when several threads try to steal tasks from the closest thread holding the lowest thread id.

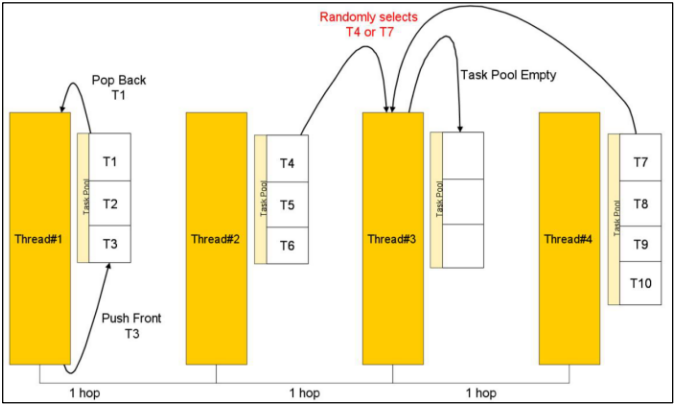
Figure 12 DFWSRPT tasks scheduler

## C. Performance evaluation of task schedulers

In order to analyze effects of our implemented task scheduling strategies, we studied the behavior of the benchmarks using these schedulers combined with priority threads allocation feature introduced in Section IV. We used for our tests the X4600 system and we conducted our tests on the different benchmarks. The majority of the benchmarks showed an improvement in performance. In this paper, FFT, Strassen and Sort benchmarks results are studied. We chose these three applications because they turned out to have the best improvements. This is due to the fact that they deal with large number of tasks and they require high amount of memory. Therefore, they require NUMA aware task schedulers to improve their performance. Following, we show the results of our schedulers.

### 1) FFT

In Figure 13 we show FFT speedup when using work-first scheduler combined to NUMA aware threads allocation and we compare it to its speedup using the two proposed task scheduling strategies, also combined to the previously introduced threads allocation technique.

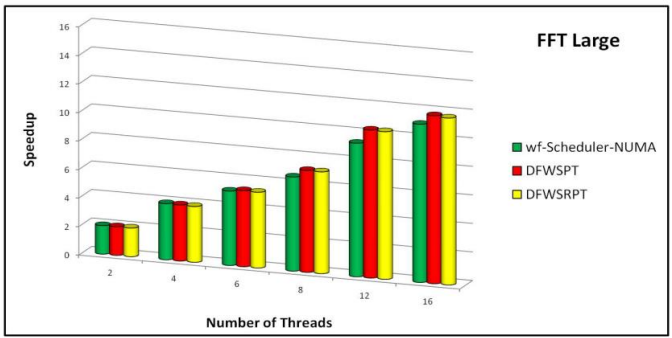
Figure 13 FFT Speedup with NUMA aware task schedulers

Results show that DFWSPT improved performance of FFT and allowed the benchmark to reach a speedup of 11.78x on sixteen threads. Hence, when using DFWSPT, FFT runs 5.85% faster than its execution time using work-first for sixteen active cores. DFWSRPT scheduler showed execution performance almost similar to DFWSPT. Both introduced schedulers improved FFT performance, mainly because they reduce the number of costly remote memory accesses whether these accesses aim to steal tasks or to read from and write to a memory location during application computation.

### 2) Sort

Same as FFT, both NUMA aware task schedulers brought almost similar performance improvements to Sort benchmark (see Figure 14). NUMA aware threads allocation model, combined to work-first scheduler, overcomes the other two schedulers for two and four active cores. But for six and eight cores, both DFWSPT and DFWSRPT enhanced performance of Sort and offered slightly better speedup than work-first based model. In addition, for more active cores, both introduced schedulers gained over the basic scheduler where DFWSPT obtained 6.32x speedup for sixteen cores to execute the benchmark 4.76% faster than with work-first.

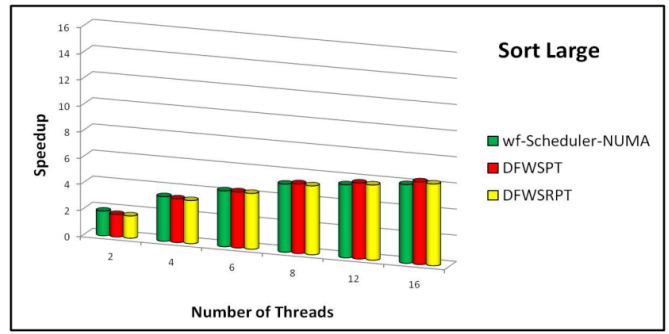
Figure 14 Sort speedup with NUMA aware task schedulers

### 3) Strassen

Like both FFT and Sort, Strassen makes good use of NUMA aware schedulers on X4600 to reduce its execution time. When four or more cores are active, both DFWSPT and DFWSRPT show better performance than work-first scheduler since remote memory accesses are significantly reduced. For Strassen, random based NUMA aware task scheduler brought better performance than DFWSPT mostly since the benchmark produces high number of task stealing between threads. For sixteen active cores, DFWSRPT can obtain a speedup of almost 12.38x over serial execution time, thus obtaining 17.03% faster execution time than with work-first based scheduler (see Figure 15).

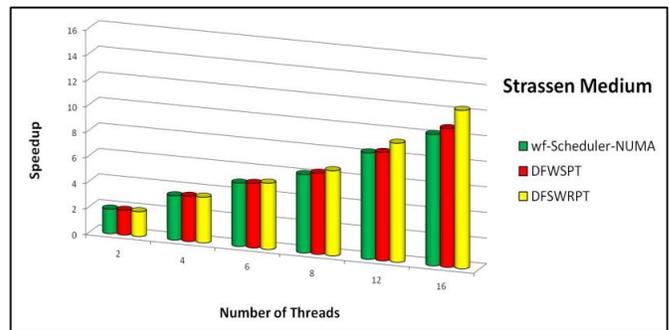
Figure 15 Strassen speedup with NUMA aware task schedulers

## VII. CONCLUSION & FUTURE WORKS

In this paper, we presented threads and tasks management models allowing Nanos runtime system to provide better efficiency for OpenMP parallel applications running on NUMA based architectures. In a first approach, we added a portable and adaptable model that generates a priority value for each core of the architecture depending on the number of close cores. Moreover, we made modifications to the runtime in order to allocate each thread's runtime system related data to the thread's attached memory. In addition, we introduced to Nanos two NUMA aware task scheduling strategies taking into consideration positions of cores when stealing tasks.

These add-ons proved good efficiency when running BOTS benchmarks on a sixteen core NUMA machine, especially for data and task intensive benchmarks. Moreover, applying this NUMA aware approach within the redundancy technique helped significantly improve performance of these benchmarks.

This work and results encourage us to analyze the performance of applications using our models on other hardware architectures. Moreover, since power consumption is an important aspect in large scale and embedded computing, building power efficient task scheduling strategies may also be considered as good candidate subject for future works.